\documentclass[twoside,11pt]{article}

\usepackage{algorithm}
\usepackage{multicol}
\usepackage{algorithmic}
\usepackage{jmlr2e}
\usepackage{setspace}

\jmlrheading{1}{2015}{1-48}{4/00}{10/00}{Qi Li}

\ShortHeadings{Distributed Parameter Map-Reduce}{Qi Li}
\firstpageno{1}

\begin{document}

\title{Distributed Parameter Map-Reduce}

\author{\name Qi Li \email liqi6@letv.com \\
       \addr Department of Big Data\\
       Leshi Internet Information $\&$ Technology Corp. \\
       Bei Jing, 100025, CHINA}

\editor{}
\maketitle

\begin{abstract}
This paper describes how to convert a machine learning problem into a series of map-reduce tasks.
We study logistic regression algorithm. In logistic regression algorithm, it is assumed that samples are independent 
and each sample is assigned a probability. Parameters are obtained by maxmizing the product of all sample probabilities.
Rapid expansion of training samples brings challenges to machine learning method. Training samples are so many that they
can be only stored in distributed file system and driven by map-reduce style programs. The main step of logistic regression
is inference. According to map-reduce spirit, each sample makes inference through a separate map procedure. But the premise 
of inference is that the map procedure holds parameters for all features in the sample. In this paper, we propose Distributed 
Parameter Map-Reduce, in which not only samples, but also parameters are distributed in nodes of distributed filesystem. 
Through a series of map-reduce tasks, we assign each sample parameters for its features, make inference for the sample and 
update paramters of the model. The above processes are excuted looply until convergence. We test the proposed algorithm 
in actual hadoop production environment. Experiments show that the acceleration of the algorithm is in linear relationship with 
the number of cluster nodes.
\end{abstract}


\section{Introduction}

 Internet companies now collect a large number of user logs every day. How to explore user's interest from these logs, so as to 
provide a personalized service has become the focal point of major internet companies attracting customers and increasing revenue. 
But exploring this treasure is not an easy task. Storing these logs requires number of machines. Analysis of these logs requires a 
lot of processors work in parallel. In recent years, hadoop platform is adopted by more and more companies. Hdfs\citep{Hdfs} provides a high 
reliable distributed filesystem. Lot of log files are divided into many small data blocks, which stored in hdfs nodes. Map-Reduce\citep{MapReduce} runs 
hdfs, it provides a simple and efficient concurrency framework calling each node's processors to solve the same problem.

  As the name suggests, map-reduce is divided into map stage and reduce stage. In map phase, it seeks subtasks are independent. 
Ideally, for each data block, a separate subtask is started without the need to interact with oter subtasks. In typical machine 
learning methods, we assume independence of samples. So it seems that a separate subtask can be started to process each sample 
block. However, samples share same parameter space and sample independence is only established in the condition of paramters determined. 
In traditional large-scale machine learning research, parameter space is in centralized storage. Subtasks on each sample block query 
paramters for its own sample from centralized parameter storage and then make inference independently. DistBelief\citep{DistBelief} divides parameter 
space into many parameter servers, and each sample server is responsible for a part of parameter space. Samples are stored in large-scale 
clusters. Each cluster node read parameters from parameter servers, compute parameter gradient each stored sampler and update parameter 
servers using these gradients. They proposed two different parameter update stratergy. Downpour SGD, each machine in the cluster 
interacts only with it's sample block involved parameter servers, reading current parameter values or update the parameters. The parameter 
updating of each sample server is independent. Sandbluster L-BFGS, storage of parameter and sample is same with Downpour SGD, and parameter 
updating of different servers is controlled by a unified parameter coodinator. Parameter gradient generated by each machine in the cluster is 
sent to parameter servers or aggregated to parameter servers through a tree structure, but does not update the parameter server immediately. 
After all the machines have generate the gradients, parameter are updated uniformly. For the parameter server, there have been a lot of research. 
For example, YahooLDA\citep{YahooLDA} implemented a dedicated server with user-definable update primitives(set, get, update) and a more principled load distribution 
algorithm. Petumn\citep{Petumn}, Graphlab\citep{Graphlab} devote to a more efficient parameter synchronization mechanism. In ~\citep{MuLi}, they develop third-generation 
parameter server. Their main contributions are: 1, support a variety of algorithms, such as sparse logistic regression, LDA. 2, the asynchronous 
communication model does not block computation. 3, the globally shared parameters are represented as vector or matrices to facilitate development  
of machine learning application. 4, related consistency hides synchronization cost and latency. They allow the algorithm designer to balance 
algorithmic convergence rate and system efficiency. 5, scalability fault tolerance and ease of use.

  Reduce phase, results generated by subtasks in the map phase are aggregated. Records with same key will be aggrated one final result. Reduce process 
is particularly suitable for commutative, associative operations, such as addition operation. Plus a large number of items, first plus part of the items, 
the intermediate result are then added with other items. It does not change the final result. Therefore, we can distribute large amounts of addition 
operations into nodes of cluster, intermediate results generated by each node are then added to produce the final result. \citep{Chu} pointed out that 
a large number of machine learning algorithms belong to Statistical Query Models. These algorithms can be written into a specific summation form. 
Therefore, they proposes multicore map-reduce framework. A map-reduce engine is responsible for spliting the samples into multiple parts. It runs a 
master, which transfer samples to different mappers, collect intermedicate results from each mapper and activate a reducer to aggrate these 
intermedicate results.

  This paper extends \citep{Chu}'s work from a single multicore machine to a distributed cluster. Unlike Statistical Query Models, our study includes only 
logistic regression algorithm. We use gradient ascent algorithm to optimize parameters. we assume these samples are independent, making it suitable for 
map process. As we pointed out earlier, different samples can only be independent in the case of the parameters determined. Each mapper must obtain the 
parameters contained in its sample block. In a distributed cluster, this is a key factor affecting the performance of parallel algorithms. Unlike previous works\citep{ParameterServerTopic,Piccolo,RDBMS}, we don't use special parameter servers to store parameter space. But like samples, parameter space is dispersed into nodes of the cluster. In parameter server mode, each mapper queries parameters from parameter server initiatively. But in our approach, each mapper obtain parameters passively. Because samples and parameters are dispersed in the sample cluster, a separate map-reduce process is started to assign each sample the current values of its parameters. The output of this map-reduce process is that each line is a sample, but these 'samples' are different from usual samples. The usual samples contain only features and feature counts. But these 'samples', contain not only features and feature counts, but also the current values of its parameters. So we call these 'samples' sufficient samples. Different sufficient samples are independent. For each sufficient sample block, start a mapper to compute parameter gradients generated 
by the sufficient sample block. We update parameter values by maximum likelihood estimation method. Therefore, an update, each parameter increment is summation of parameter gradients produced by different samples. Thereby parameter increment calculation is applicable for reduce process. After obtaining the parameter increments, start a separate map-reduce process to generate new parameter values based on old parameter values and increments. These new parameter values are assigned to each sample and new sufficient samples are produced.

 Later sections in this article are organized as follows. In the second chapter, we briefly the logistic regression model, including variables, parameters, 
objective function and inference process. Logistic regression is applicable to the Distributed Parameter Map-Reduce Framework we propose. In third chapter, 
we describe  Distributed Parameter Map-Reduce method in detail. Through a series of map-reduce tasks, we build parameter invert index, generate sufficient sample, 
compute gradient and update parameter. In this chapter, we describe each map-reduce process in detail. To get sufficient samples, we start a map-reduce process 
to build parameter invert index. Every line in the output is 'parameter$\rightarrow$sample' relationship. If a parameter occurs in large amount of samples, the line will be very long and take up a lot of bytes, making sample blocks containing invert index uneven. Some parameter invert index takes several data blocks, and some 
data block contains lots of invert indexs. This also affects the map-reduce process generating sufficient samples. Some mapper will take a very long time fo finish. 
In fourth chapter, we introduce high-frequency parameter sharding, which produces invert index in uniform distribution. In the fifth chapter, we use a series of 
map-reduce tasks to make prediction for test samples. The procedure is similar to one training iteration. Unlike training process, we don't compute gradients for each sample, but make prediction for each sample. In the sixth chapter, we introduce the experimental design and results. In this chapter, we mainly validate 
the acceleration effect of Distributed  Parameter Map-Reduce. Our experiments show that the accelaration proportion of Distributed Parameter Map-Reduce and the 
number of cluster nodes is the kind of linear relationship. Therefore, Distributed Parameter Map-Reduce is especially suitable for large-scale machine learning 
in distributed cluster.
  
\section{Logistic Regression}
 \label{sec:logistic}
  Logistic regression is a typical regression algorithm. In the following, $x^{(i)}$ represents a document, $y^{(i)}$ represents the label. Assume samples are $(x^{(1)},y^{(1)}),\cdots,(x^{(D)},y^{(D)}),\;y^{(i)}\in\{0,1\}$, $\theta$ represents a vector of all parameters, $\theta_j$ represents parameter corresponding to feature $j$.
Each sample $x^{(i)}$'s probability is expressed as a sigmoid function. The probability that $x^{(i)}$ is assigned label $1$ given $\theta^{(i)}$ ($\theta^{(i)}$ represents parameter coressponding to $x^{(i)}$) is equal to
\begin{eqnarray}
p(y^{(i)}=1|x^{(i)};\theta^{(i)})&=&h_{\theta^{(i)}}(x^{(i)})\\\nonumber
         &=&\frac{1}{1+exp(-{\theta^{(i)}}^Tx)}\\\nonumber
         &=&\sigma({\theta^{(i)}}^Tx)\nonumber
\end{eqnarray}
We use maximum likelihood estimation to solve the optimal parameters, so the objective function is 
\begin{equation}
L(Y,X;\theta)=\prod_{i=1}^D p(y^{(i)}|x^{(i)},\theta^{(i)})
\end{equation} 
This is equal to minimizing the following cost function
\begin{eqnarray}
J(\theta)&=&-\sum_{i=1}^D \{ y^{(i)}log(h_{\theta^{(i)}}(x^{(i)}))\\\nonumber
         &&+(1-y^{(i)}log(1-h_{\theta^{(i)}}(x^{(i)}))\}\nonumber
\end{eqnarray}

  The optimal solution of $J(\theta)$ can not be solved directly. We use optimization method\citep{Optimization} such as gradient descent to approach 
the optimal solution. The most important is to calculate the gradients:
\begin{equation}
\bigtriangledown_jJ(\theta)=\sum_{i=1}^Dx_j^{(i)}(h_{\theta^{(i)}}(x^{(i)})-y^{(i)})
\end{equation}
In $\bigtriangledown_jJ(\theta)$, the key is calculation of $h_{\theta^{(i)}}$. We call such probability calculation the inference of the sample. It is 
noteworthy that these probabilities are calculated independently given parameters $\theta^{(i)}$ for sample $x^{(i)}$.

 If we use gradient descent method to approximate the optimal solution. In each iteration, the parameters are updated using the following formula
\begin{equation}
\label{grad}
\theta_j=\theta_j-\alpha\cdot\bigtriangledown_jJ(\theta)
\end{equation}
$\alpha$ is learning rate, which is a postive constant. In each iteration, increment for parameter $\theta_j$ is $\bigtriangledown_jJ(\theta)$ multiplied by 
a constant. We substitute $\bigtriangledown_jJ(\theta)$ into formula~\ref{grad}
\begin{equation}
\theta_j=\theta_j+\alpha\sum_{i=1}^D x_j^{(i)}(y_j^{(i)}-h_{\theta^{(i)}})
\end{equation}
As can be seen from the above equation, updating for parameter $\theta_j$ is summation form. In the above discussion, we pointed out that parameter 
updating of this form applies to reduce process. So in mappers, $x_j^{(i)}(y_j^{(i)}-h_{\theta^{(i)}})$ is calculated for each sample. The output 
is aggregated through reduce process into final increment for parameter $\theta_j$.

\section{Distributed Parameter Map-Reduce}
 A machine learning task can be converted into a series map-reduce tasks. In this chapter, we describe each map-reduce task in detail. 
For ease of discussion, we first determine parameter granularity. In logistic regression, one feature corresponds to a parameter, which 
reflects the feature's weight reallocation between two labels. In the following discussion, we store parameter space by feature. Each 
row represents a key-value, with feature as key and parameter corresponding to the feature as value.

 Before discussing each map-reduce task, we first introduce the overall structure of Distributed Parameter Map-Reduce. This helps to better 
understand the role of each map-reduce task. Algorithm~\ref{alg:main} describes the structure of our algorithm. 

\begin{algorithm}[tb]
   \caption{Distributed Parameter Map-Reduce}
   \label{alg:main}
{\bfseries Signature dpmr(trainInput, paraValueOutput)}
\begin{algorithmic}[1]
   \STATE initParameters(trainInput, paraValueOutput)
   \STATE invertDocuments(trainInput, docInvertOutput)
   \FOR{$t=1$ {\bfseries to} $Iter$}
   \STATE distributeParameters(paraValueOutput, docInvertOutput, paraDistributeOutput)
   \STATE restoreDocuments(paraDistributeOutput, docRestoreOutput)
   \STATE computeGradients(docRestoreOutput, gradComputeOutput)
   \STATE updateParameters(paraValueOutput, gradComputeOutput, paraUpdateOutput)
   \STATE copy(paraUpdateOutput, paraValueOutput)
   \ENDFOR
\end{algorithmic}
\end{algorithm}

 The entrance of Algorithm~\ref{alg:main} is dpmr(trainInput,paraValueOutput). Enter the training corpus and output parameters that 
minimizes cost function. Input, output and intermediate results are stored in the form of hdfs files. There are mainly six map-reduce tasks: 
initParameters, invertDocuments, distributeParameters, restoreDocuments, computeGradients, updateParameters. First, extract all features 
from the corpus and initialize parameter for each feature. The output is in the form of 'feature->parameter'. InvertDocuments inverts each 
sample by feature, and build index with the form of 'feature$\rightarrow$sample'. After the completion of initParameters and invertDocuments, for each 
feature, we can get both the corresponding parameter value and all samples in which the feature appears. DistributeParameters combines the 
two. For each sample, we can get both the feature and current parameter value corresponding to the feature. But each line of paraDistributeOutput 
contains only one feature. RestoreDocuments brings together all features contained in one sample. So far, for each sample, we know all the 
features and current parameter for each feature. We call each row of sample in docRestoreOutput 'sufficient sample'. ComputeGradients makes 
inference for each 'sufficient sample' independently and compute parameter gradient for each feature. For any feature, there is both current 
value and gradient of its parameter. So optimization methods can be called to update current value of the parameter. New parameter vlues replace 
the current and the copy method update parameters to the latest values. The copy method is hdfs's own function, such as copy method of FileUtilClass. 
In addition, the algorithm must have the termination condition, determining whether the parameter value is already optimal. This can be done 
by adding one step  between step 5 and step 6 in Algorithm~\ref{alg:main} to calculate the current value of the objective functions. This step is 
to start a separate map-reduce task like computeGradients to complete. Record history values of the objective function to determine whether 
the current iteration can be terminated. Later in this section, we describe six map-reduce processes in detail.

\subsection{initParameters}

\begin{algorithm}[tb]
   \caption{initParameters(trainInput, paraValueOutput)}
   \label{alg:initPara}
\begin{multicols}{2}
{\bfseries mapper($longId$, $S^{(i)}$)}
\begin{algorithmic}[1]
   \FOR{each $token$ in $S^{(i)}$}
   \STATE $f=token\rightarrow f$
   \STATE $emit(f, 1)$
   \ENDFOR
\end{algorithmic}

{\bfseries reducer($f$, $iterator$)}
\begin{algorithmic}[1]
   \STATE static\{
   \STATE   \hspace{0.3cm}$para=0$
   \STATE   \hspace{0.3cm}$type='p'$
   \STATE   \hspace{0.3cm}$paraStr$=$concat$($'$\hspace{0.3cm}$'$, $type$, $para$)
   \STATE \}
   \STATE $emit(f,paraStr)$
\end{algorithmic}
\end{multicols}
\end{algorithm}
Algorithm~\ref{alg:initPara} describes the process of initParameters. InitParameters extracts all the features from training corpus, and initialize 
the parameter of each feature. Training data format is that each line stores one sample $S^{(i)}$, and each sample stores its features in feature 
hash form. $S^{(i)}$ contains label part and feature part. Label part stores annotation information of $S^{(i)}$. Label space is generally much 
smaller with respect to feature space. We create dictionary for the label space. In $S^{(i)}$, the label is an integer that represents the label 
index in the dictionary. In logistic regression, there are two different categories. In each sample $S^{(i)}$, we use an integer of 0 or 1 to 
indicate the actual category $S^{(i)}$ belongs.

  The feature part is a series of tokens. Each token corresponds to a feature and has the form of '$f:count$'. Because feature space is generally large 
and easy to change, we don't build dictionary for feature space. $f$ is a string that uniquely identifies the feature and count represents the number 
of times $f$ appears in $S^{(i)}$. 

  The input of mapper is sample $S^{(i)}$ and $longId$ automatically generated by hadoop. The mapper only processes the feature part of $S^{(i)}$. For 
each token, output a record with $f$ as key, $1$ or $null$ as value.

  The input of reducer is feature $f$ and all the values corresponding to key $f$. These values are aggregated into a iterator. In initParameters, 
our goal is extracting feature $f$, not focus on values in iterator. For feature $f$, we must intitialize the corresponding parameter. Every reduce 
task that hadoop started make the same initialization for all features. In static code of reduce class, the parameter value is initialized to $0$. 
For subsequent processing, $type$ field is added to indicate thate the output of initParameters is parameter information. In output of reducer, key 
is $f$ and value is $paraStr$. $paraStr$ contains both the $type$ field and the initialization parameter information.

\subsection{invertDocuments} 
 The key step of Distributed Parameter Map-Reduce is to build 'index' of 'feature$\rightarrow$sample' form. Algorithm~\ref{alg:docInvert} describes 
the detail process of building 'index'. This 'index' is not same with usual index. It is only plain composed of lines of 'key$\rightarrow$value' pairs and 
stored in hdfs. With this index, for one feature, the corresponding parameter and the sample list which it appears can be sent to the same reducer. 
In the reducer, sample get features it contains and current parameter values corresponding to its features.

\begin{algorithm}[tb]
   \caption{invertDocuments(trainInput, docInvertOutput)}
   \label{alg:docInvert}
\begin{multicols}{2}
{\bfseries mapper($longId$, $S^{(i)}$)}
\begin{algorithmic}[1]
   \STATE $docId=generateRandomId(S^{(i)})$
   \STATE $label=S^{(i)}\rightarrow label$
   \STATE $type='i'$
   \STATE $num=1$
   \FOR{each $token$ in $S^{(i)}$}
   \STATE $f=token\rightarrow f$
   \STATE $count=token\rightarrow count$ 
   \STATE $units=concat(':',docId,count,label)$
   \STATE $value$=$concat$($'$\hspace{0.3cm}$'$, $type$,$num$, $units$)
   \STATE $emit(f, value)$
   \ENDFOR
\end{algorithmic}

{\bfseries combiner(reducer)($f$, $iterator$)}
\begin{algorithmic}[1]
   \STATE  $sum=0$
   \STATE  $docUnits=''$
   \STATE  $type='i'$
   \FOR{each $value$ in $iterator$}
   \STATE $sum+=value\rightarrow num$
   \STATE $units=value\rightarrow units$ 
   \STATE $docUnits$=$concat$($'$\hspace{0.3cm}$'$,$docUnits$,$units$)
   \ENDFOR
   \STATE $value$=$concat$($'$\hspace{0.3cm}$'$, $type$,$sum$, $docUnits$)
   \STATE $emit(f,value)$
\end{algorithmic}
\end{multicols}
\end{algorithm}

\begin{algorithm}[tb]
   \caption{distributeParamters(paraValueOutput,docInvertOutput,paraDistributeOutput)}
   \label{alg:paraDistribute}
\begin{multicols}{2}
{\bfseries mapper($longId$, $paraInfo$)}
\begin{algorithmic}[1]
   \STATE $f=paraInfo\rightarrow f$
   \STATE $value=paraInfo\rightarrow value$
   \STATE $emit(f,value)$
\end{algorithmic}

{\bfseries reducer($f$, $iterator$)}
\begin{algorithmic}[1]
   \STATE  $para=''$
   \STATE  $docUnits=''$
   \FOR{each $value$ in $iterator$}
   \STATE $type=value\rightarrow type$
   \IF{$type=='i'$}
   \STATE $docUnits=value\rightarrow docUnits$
   \ENDIF
   \IF{$type=='p'$}
   \STATE $para=value\rightarrow para$
   \ENDIF
   \ENDFOR
   \FOR{each $unit$ in $docUnits$}
   \STATE $docId=unit\rightarrow docId$
   \STATE $count=unit\rightarrow count$
   \STATE $label=unit\rightarrow label$
   \STATE $value=concat(':',label,f,count,para)$
   \STATE $emit(docId,value)$
   \ENDFOR
\end{algorithmic}
\end{multicols}
\end{algorithm}

  The input of Algorithm~\ref{alg:docInvert} is also training corpus. In order to build inverted index, we first assign each sample a 
random Id. The granularity of inverted index is feature $f$, so the key of output is $f$. Value is composed of units. Each unit 
corresponding to one sample. In output of mapper, each value has only one unit. Reducer gathers together all the values corresponding to 
feature $f$. Every unit is composed of $f$,\hspace{0.1cm}$count$,\hspace{0.1cm}$label$. So we can recover the sample later. In addition, type field is added to 
represent that the output record is sample inverted index record. Reducer process and combiner process is same. Some feature $f$ is 
more common, so its sample list is very long. This will affect the parallel performance of the algorithm. The $count$ statistics is to 
facilitate the subsequent sharding of feature $f$ with long sample list. The key $f$ is subdivided into a series of sub-keys. Each 
sub-key corresponds to part of sample list for feature $f$. We currently do not make in depth description of the sharding process. 
It will discussed in detail in the next section.

\subsection{distributeParameters and restoreDocuments}
Algorithm~\ref{alg:initPara} initialize the parameter of the feature. Algorithm~\ref{alg:docInvert} build 'feature$\rightarrow$sample' index. 
The output keys of two algorithms are both feature $f$, which provides the basis for starting the map-reduce process in Algorithm~\ref{alg:paraDistribute}. 
Parameter and sample list with same key are sent to the same reducer. Reducer determines whether the record is parameter or sample list 
according to the $type$ field. From above, we know that, in sample list $docUnits$, each $unit$ corresponds to one sample. For each $unit$, 
reducer will output one record. So $docId$ and $f$ swap positions. The $docId$ is key and $f$ is part of vlue. This is similar to the process of 
recovering sample from 'feature$\rightarrow$sample' index. But this is not simple recovery, in which parameter corresponding to the feature 
is also included. From one 'feature$\rightarrow$sample' record, we can only recover one feature for the sample. Algorithm~\ref{alg:docRestore} gathers 
together all the features corresponding to same $docId$, resulting in the complete 'sample' for this $docId$.

  We call each sample in the output of Algorithm~\ref{alg:docRestore} sufficient sample. The sample contains not only its features ($f$ and 
$count$ of $f$) and $label$ information., but also current values of parameters corresponding to the features that is necessary to make 
inference for the sample. The advantage of sufficient sample is that it can be made inference independent of others sufficient samples.

\begin{algorithm}[tb]
   \caption{restoreDocuments(paraDistributeOutput, docRestoreOutput)}
   \label{alg:docRestore}
\begin{multicols}{2}
{\bfseries mapper($longId$, $docPartInfo$)}
\begin{algorithmic}[1]
   \STATE $docId=docPartInfo\rightarrow docId$
   \STATE $value=docPartInfo\rightarrow value$
   \STATE $emit(docId, value)$
\end{algorithmic}

{\bfseries combiner(reducer)($f$, $iterator$)}
\begin{algorithmic}[1]
   \STATE  $values=''$
   \STATE  $docUnits=''$
   \STATE  $type='i'$
   \FOR{each $value$ in $iterator$}
   \STATE $values$=$concat$($'$\hspace{0.3cm}$'$,$values$,$value$)
   \ENDFOR
   \STATE $emit(docId,values)$
\end{algorithmic}
\end{multicols}
\end{algorithm}

\subsection{computeGradients}
  From Section~\ref{sec:logistic}, we know that, in logistic regression the parameter gradient can be expressed as the summation form. Algorithm~\ref{alg:docRestore} 
output all the sufficient samples. The mapper of Algorithm~\ref{alg:gradCompute} makes inference independently for each sample. The minimum granularity of parameter 
space in logistic regression is feature. Each feature corresponds to one parameter. $grad$, $prob$, $empir$ respectively store the gradient, inference probability, 
experience value. In logistic regression, we make inference for each parameter and calculate its posterior probability. Gradient is equal to the expectation value 
minus experience value. Reducer adds up all the parameter gradient with same key, as the final parameter gradient for the feature.
\begin{algorithm}[tb]
   \caption{computeGradients(docRestoreOutput, gradComputeOutput)}
   \label{alg:gradCompute}
\begin{multicols}{2}
{\bfseries mapper($longId$, $Suff^{(i)}$)}
\begin{algorithmic}[1]
   \STATE $grad=0$
   \STATE $prob=0$
   \STATE $empir=0$
   \STATE $type='q'$
   \STATE $label=Suff^{(i)}\rightarrow label$
   \STATE $inf=inference(Suff^{(i)},label)$
   \STATE $emp=empirCount(Suff^{(i)},label)$
   \FOR{each $value$ in $iterator$}
   \STATE $empir=emp[f]$
   \STATE $prob=inf[f]$
   \STATE $grad=func(empir,prob,label)$
   \STATE $value$=$concat$($'$\hspace{0.3cm}$'$,$type$,$grad$)
   \STATE $emit(f,value)$
   \STATE $values$
   \ENDFOR
\end{algorithmic}

{\bfseries combiner(reducer)($f$, $iterator$)}
\begin{algorithmic}[1]
   \STATE  $summation=0$
   \STATE  $type='q'$
   \STATE  $grad=0$
   \FOR{each $value$ in $iterator$}
   \STATE $grad=value\rightarrow grad$
   \STATE $summation+=grad$
   \ENDFOR
   \STATE $value$=$concat$($'$\hspace{0.3cm}$'$,$type$,$summation$)
   \STATE $emit(f,value)$
\end{algorithmic}
\end{multicols}
\end{algorithm}
\subsection{updateParameters} 

 The final setp is updating parameter. Algorithm~\ref{alg:paraUpdate} describes the process of updating parameter in detail. ParaValueOutput stores current value of 
the parameter. GradComputeOutput is output of Algorithm~\ref{alg:gradCompute}, and stores the parameter gradient. Use gradient descent method to update the parameter. $\theta_j=\theta_j-\alpha\cdot \bigtriangledown_jJ(\theta)$, where $\alpha$ is the learning rate. According, step (12) in Algorithm~\ref{alg:paraUpdate} 
ia actually $newPara=para-\alpha\cdot grad$. $para$ and $grad$ respectively store current parameter and gradient value corresponding to feature $f$. $newPara$ is 
new parameter corresponding to feature $f$.
.

\begin{algorithm}[tb]
   \caption{updateParameters(paraValueOutput,gradComputeOutput,paraUpdateOutput)}
   \label{alg:paraUpdate}
\begin{multicols}{2}
{\bfseries mapper($longId$, $paraInfo$)}
\begin{algorithmic}[1]
   \STATE $f=paraInfo\rightarrow f$
   \STATE $value=paraInfo\rightarrow value$
   \STATE $emit(f,value)$
\end{algorithmic}

{\bfseries reducer($f$, $iterator$)}
\begin{algorithmic}[1]
   \STATE  $para=0$
   \STATE  $grad=0$
   \FOR{each $value$ in $iterator$}
   \STATE $type=value\rightarrow type$
   \IF{$type=='q'$}
   \STATE $grad=value\rightarrow grad$
   \ENDIF
   \IF{$type=='p'$}
   \STATE $para=value\rightarrow para$
   \ENDIF
   \ENDFOR
   \STATE $newPara$=$optimize(para,grad)$
   \STATE $type='p'$
   \STATE $value$=$concat$($'$\hspace{0.3cm}$'$,$type$,$newPara$)
   \STATE $emit(f,value)$
\end{algorithmic}
\end{multicols}
\end{algorithm}
\section{Sharding}
   
   According to zip's law, $freq{f}\propto 1/rank(f)$, $freq(f)$ is the frequency of feature $f$. If feature f is stored in descending frequency, $rank(f)$ 
is the position of the feature $f$ in the sort results. This means that the frequency distribution of features is very uneven, with a large number of 
features having a smaller frequency and small number of features having a greater frequency. A small amount of high-frequency features will seriously 
affect the performance of Distributed Parameter Map-Reduce algorithm. First, in 'feature$\rightarrow$sample' index, if the frequency of one feature is 
very large, the corresponding sample list will be very large. Sample list corresponding to one feature is stored in the same line in 'feature$\rightarrow$sample' 
index. If the sample list is too large, the line will be too long. In Algorithm~\ref{alg:docInvert}, a sample is assigned a random string type $docId$. Suppose 
$docId$ is composed of ten random letters or numbers and the sample list size is 100 million, the line will occupy at least 1G of storage space. In 
hdfs, block size is generally set to 64M, and the line will take up about 20 blocks. The block distribution of each line storage space in 'feature$\rightarrow$sample' index will be very uneven. Secondly, a few features with large size sample list make mappers or reducers processing these features much slower than mappers or reducers processing large amount of low-frequency features, which affectiong the overall parallel effect of Distributed Parameter Map-Reduce algorithm.

\subsection{sharding process}

\begin{algorithm}[tb]
   \caption{Sharding Distributed Parameter Map-Reduce}
   \label{alg:shardingmain}
{\bfseries Signature dpmr\_sharding(trainInput,\hspace{0.2cm}paraValueOutput)}
\begin{algorithmic}[1]
   \STATE initParameters(trainInput,\hspace{0.2cm}paraValueOutput)
   \STATE invertDocumentsSharding(trainInput,\hspace{0.2cm}docInvertShardOutput)
   \STATE invertParameters(docInvertShardOutput,\hspace{0.2cm}paraInvertOutput)
   \FOR{$t=1$ {\bfseries to} $Iter$}
   \STATE distributeParametersSharding(paraValueOutput,\hspace{0.2cm}paraInvertOutput,\hspace{0.2cm}paraDistributeShardOutput)
   \STATE distributeParameters(paraDistributeShardOutput,\hspace{0.2cm}docInvertShardOutput,\hspace{0.2cm}paraDistributeOutput)
   \STATE restoreDocuments(paraDistributeOutput,\hspace{0.2cm}docRestoreOutput)
   \STATE computeGradientsSharding(docRestoreOutput,\hspace{0.2cm}gradComputeShardOutput)
   \STATE updateParameters(paraValueOutput,\hspace{0.2cm}gradComputeShardOutput,\hspace{0.2cm}paraUpdateOutput)
   \STATE copy(paraUpdateOutput,\hspace{0.2cm}paraValueOutput)
   \ENDFOR
\end{algorithmic}
\end{algorithm}

  To solve this problem, we make sharding for small amount of high-frequency features. The sharding feature $f$ is as follows: suppose complete or part count of 
feature $f$ is $c$, feature $f$ is divided into $N$ sub-features, each of which is of the form '$i\_N|f,1\le i \le N$'. Sample list of feature $f$ is assigned to $N$ 
sample lists corresponding to $N$ sub-features. In 'feature$\rightarrow$sample' index, $N$ sub-features substitute the parent feature $f$. So a long line is divided 
into $N$ short lines. These $N$ sub-features will be assigned to $N$ mappers or reducers in the subsequent processing. Sharding process can be nested, namely 
sub-feature $f_s$ is then to be made sharding.

  If sharding, the entrance of Distributed Parameter Map-Reduce must make appropriate adjustments, as shown in Algorithm~\ref{alg:shardingmain}. InvertDocumentSharding is similar to Algorithm~\ref{alg:docInvert}, but its mapper,combiner and reducer make feature sharding. Enter one feature $f$, 
output $N$ sub-features $f_s$. In combiner,reducer, count $c$ of feature $f$ is equal to the size of $f$'s sample list after aggregation of combiner or reducer. 
Sharding in mapper needs external incoming feature frequency statistics. The number of high-frequency is smaller. Make statistics in advance and pass statistics 
results of high-frequency into mapper. This is feasible. Of course, sharding in mapper is not necessary. This can be done only in combiner and reducer. So that 
no external feature frequency is needed. Sharding in mapper, combiner, reducer is performed sequentially nested.
 
  InvertDocuments build 'parent feature $\rightarrow$ sub feature' index for sub features after sharding and their parent features. All sub features corresponding to 
parent feature $f$ are stored in the same line. Suppose that sub feature $f_s$ is $i\_N|f$, then its parent feature is $f$. In paraInvertOutput, feature $f$ corresponds to one line, in which key is $f$ and string representation of ${f_s|f_s\in sub(f)}$ is value. DistributeParametersSharding sets parameter of sub 
feature $f_s$ as parameter of its parent feature $f$. Therefore, all sub features corresponding to one parent feature have the same parameter. Parameter of 
sub feature $f_s$ in paraDistributeOutput is input into distributeParameters in place of parameter of parent feature $f$. The other processes are basically 
same as Algorithm~\ref{alg:main}. The only change is that computeGradients make a slight adjustment. In step (13) of Algorithm~\ref{alg:gradCompute}'s mapper, 
parent feature $f$ is extracted from sub feature $f_s$. The output is change from $emit(f_s,value)$ into $emit(f,value)$. Pseudocode of invertDocumentSharding, 
invertParameters, distributeParametersSharding respectly corresponds to Algorithm~\ref{alg:docInvertShard}, Algorithm~\ref{alg:paraSub}, Algorithm~\ref{alg:paraSubDis} in the appendix. 

\subsection{effective sharding}
 \begin{algorithm}[tb]
   \caption{Classifying Distributed Parameter Map-Reduce}
   \label{alg:classifyingmain}
{\bfseries Signature dpmr\_classifying(testInput,\hspace{0.2cm}testOutput)}
\begin{algorithmic}[1]
   \STATE invertDocumentsSharding(testInput, docInvertShardTestOutput)
   \STATE invertParameters(docInvertShardTestOutput, paraInvertTestOutput)
   \STATE distributeParametersSharding(paraValueOutput,\hspace{0.2cm}paraInvertTestOutput,\hspace{0.2cm}paraDistributeShardTestOutput)
   \STATE distributeParameters(paraDistributeShardTestOutput,\hspace{0.2cm}docInvertShardTestOutput,\hspace{0.2cm}paraDistributeTestOutput)
   \STATE restoreDocuments(paraDistributeTestOutput, docRestoreTestOutput)
   \STATE logisticTest(docRestoreTestOutput, testOutput)
\end{algorithmic}
\end{algorithm}
  Carefully designing of sharding function in Algorithm~\ref{alg:docInvertShard} can effectively reduce network data transmission. We can start from the 
following two aspects. 1, try to make sharding key $f_s$ and the corresponding samples are assigned to the same reducer. So output of step (17) in Algorithm~\ref{alg:paraDistribute} is stored in local data blocks. This method is effective for high-frequency features, but has little effect on the low-frequency features. 
2, in the case which the first method is not suitable, such as low-frequency features, try to make sharding key $f_s$ and the corresponding samples are 
allocated on machines on the same rack. This can reduce data transmission between different racks, to improve transmission efficiency. For Algorithm~\ref{alg:paraDistribute}, even the entire Distribute Parameter Map-Reduce algorithm, an issue which should be also noted is the inherent characteristic of hdfs. 
The same data block has multiple backups. Multi-backup inevitably brings more network data transmission. We should consider parallelling effect of the algorithm, 
network data transmission, system fault tolerance and other factors to determine the number of backups. Maybe different backup strategies are adopted in 
different stages. For example, use single backup for Algorithm~\ref{alg:paraDistribute}.

\section{Classifying}
 The whole process is similar to the training process Algorithm~\ref{alg:shardingmain}. Format of testInput is same with that of trainInput. Each line 
has a test sample. Each row of testOutput stores a prediction result. The format is $line.=.<docId> <sep> <example\_label> <sep> <predict\_label>$. 
Test corpus and model parameters( paraValueOutput) are both stored in hdfs. Dpmr\_classifying uses a series of map-reduce procedures to finish prediction. 
The result testOutput is also stored in the form of hdfs file. The only new method is logisticTest. The procedure is similar to Algorithm~\ref{alg:gradCompute}. 
But it has no reduce stage, only map stage. For each line of sufficient sample $x^{(i)}$, map stage of Algorithm~\ref{alg:gradCompute} outputs the corresponding parameter gradient values. But for each line of sufficient sample $x^{(i)}$, logisticTest outputs the probability $p(y^{(i)}=1|\theta^{(i)},x^{(i)})$.  

\section{Experiments}

\begin{table}[tb]
\caption{Running time(minutes) of various map-reduce tasks under different conditions}
\label{running-dpmr}
\vskip 0.15in
\begin{center}
\begin{small}
\begin{sc}
\begin{tabular}{lcccc}
\hline
Map-Reduce & (33,25) & (100,75) & (200,150)&loop \\
\hline
invertParametersSharding & 315& 112& 51&$\times$ \\
invertDocumets    & 189& 69& 30&$\times$  \\
distributeParametersSharding    & 237& 85&39&$\surd$  \\
distributeParameters    & 221& 78& 43&$\surd$          \\
restoreDocuments     & 175& 54& 30&$\surd$ \\
computeGradientsSharding    & 259& 99& 45&$\surd$ \\
updateParameters     & 161& 51& 27&$\surd$ \\
average of one iteration   & 1053& 367& 185& \\
\hline
\end{tabular}
\end{sc}
\end{small}
\end{center}
\vskip -0.1in
\end{table}

 In order to verify the proposed Distributed Parameter Map-Reduce algorithm, we test the following data set. The dataset sample size exceeds 2T and the feature 
size exceeds 500G. Approximately 20 billion samples and 50 billion features. Testing focuses on the following two aspects: first test running time under 
different resources( number of mappers, reducers). The second is the convergence speed. Acceleration proportion of the algorithm is in linearly proportional 
relationship with the number of nodes( mappers, reducers). After a few iterations, the algorithm will converge. In our experiments, after two iterations, the 
algorithm basically reaches the level of convergence. 

\subsection{acceleration proportion}
  
\begin{figure}[t]
\vskip 0.2in
\begin{center}
\centerline{\includegraphics[width=\columnwidth]{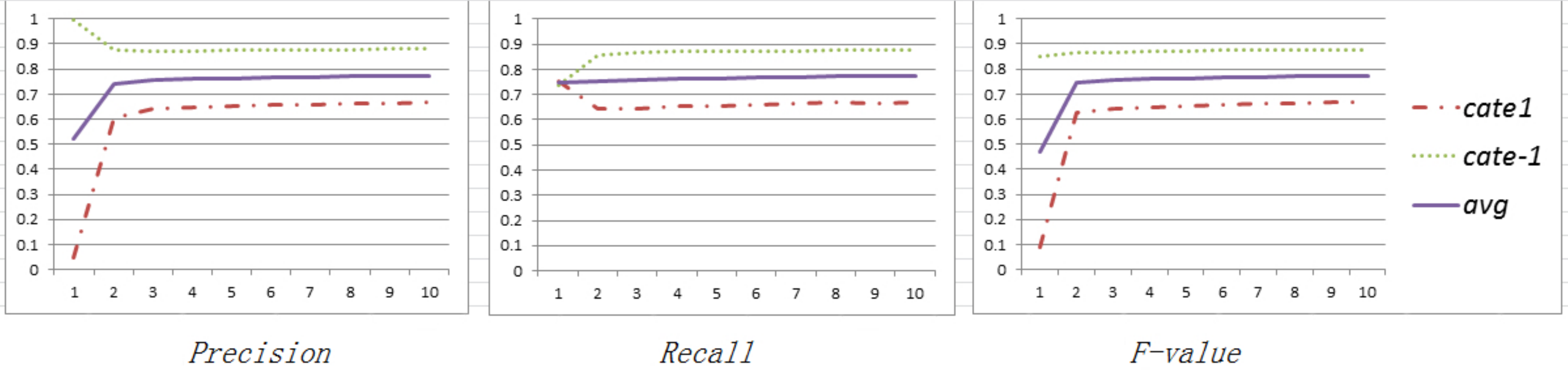}}
\caption{The convergence effect of Distributed Parameter Map-Reduce algorithm. Class $+1$, $-1$ are assessed separately. Compute respective scores, the 
 average of two as the whole score. $cate1$, $cate-1$ represent respectively $+1$, $-1$, $avg$ is the average of $cate1$ and $cate-1$. Y-axis is score, 
 X-axis is the number of iterations. The left is accurate rate, the middle is the recall rate, and the right is $F\hspace{0.1cm}value$. $Accuracy$ and $F\hspace{0.1cm}value$ bias Class $-1$ at the first iteration. At the second iteration, $Accuracy$ and $F\hspace{0.1cm}value$ of Class $+1$ reach a reasonable level. Class $-1$ declines slightly. When the first 
iteration, recalls of Class $+1$, $-1$ are almost equal. When the second iteration, Class $-1$ increases slightly and Class $+1$ decreases slightly.}
\label{convergence-dpmr}
\end{center}
\vskip -0.2in
\end{figure}

  In this experiment, we count running time in the following three resources: (1) 33 mappers, 25 reducers (2) 100 mappers, 75 reducers (3) 200 mappers, 150 reducers. 
Average running time of each map-reduce procedure of Distributed Parameter Map-Reduce algorithm and that of one iteration is shown in Table~\ref{running-dpmr}. 
Map-reduce procedure that takes less time is ignored.

 InvertParameters, distributeParametersSharding, updateParameters only process parameter space, so they all take a relatively small time. InvertDocumentsSharding, 
distributeParameters both process sample space. Two processed are just opposite. DistributeParameters takes a slightly shorter time than invertDocumentsSharding, 
because the sharding method in invertDocumentsSharding is carefully designed. But currently we only implement that sharding key $f_s$ and the corresponding 
sample list are allocated on the same reducer. The design that they are allocated on the same rack is not really realized so far. Since invertParameters only once in 
the entire Distributed Parameter Map-Reduce algorithm, but distributeParameters to be repeated, carefully designed invertDocumentSharding saves the overall 
time cost of the algorithm. Although restoreDocuments process sample space, but the input key of mapper and the output key of reducer are both $docId$. So less 
network traffic, thus faster. For computeGradientSharding, calculation process takes longer time. Network data transmission is mainly parameter space, so it takes 
a relatively short time. But computeGradientsSharding's speed is not slower than distributeParameters. (100,75) relative to (33,25), 3 times of the number of 
nodes, speedup ratio is 3.1. While (200,75) relative to (33,25), 6 times the number of nodes, speedup ratio is 6. This indicates that speedup ratio is in 
approximately linear relationship with the number of nodes. On the experiment cluster, a data block has three blocks. Compression storage of data, lzo algorithm 
is used by default.

\subsection{convergence effect}

   Since convergence effect and the number of nodes( mappers, reducers) have no relationship. We only make statistics of (200,150) case. After each iteration, 
test procedure Algorithm~\ref{alg:classifyingmain} is called. Statistical results are shown in Figure~\ref{convergence-dpmr}.

   Class +1, -1 are assessed separately. +1, -1 ratio of the sample is roughly 3:1. Compute respective scores, the average of two as the whole score. The three 
targets, Precision, Recall and F value, are computed. They correspond respectively to Figure 1, 2, 3. F value is comprehensive assessment of Precision, Recall, 
$f=2pr/(p+r)$. The figures show that, whether it is Precision, Recall or F value, after two iterations, all reach fundamental convergence level. First iteration, 
initial parameter value is 0. After the first iteration, the algorithm does not converge. However, it makes a preliminary allocation of parameter weight, to 
determine whether a parameter is postive effect or negative effect for each class. Second iteration is parameter tuning refinement process. After this iteration, 
each parameter weight substantially reaches a reasonable value.

\section{Conclusion and Future Work}

 In this paper, we propose Distributed Parameter Map-Reduce algorithm, implementing training and classifying for logistic regression through a series of 
map-reduce tasks. The input, output and intermediate results are all stored in the form of hdfs files. Through several continuous map-reduce procedures, 
we assign each sample involved parameters current values, make it a sufficient sample. Different sufficient samples are independently, applying to map procedure. 
The update of logistic regression can be expressed as summation form, applying to reduce procedure.
 
 Distributed Parameter Map-Reduce algorithm can easily process T level samples, and more importantly, at the same time, it can also process T level feature 
space. Feature space is dispersed into nodes of the hdfs. All procedures are map-reduce tasks. Increased feature space will not affect the algorithm parallelism 
and scalability. For larger sample or feature space, spread the load on more nodes. Acceleration proportion is in linear relationship with the number of nodes( mappers and reducers) of the cluster. More nodes will linearly bear more loads or reduce algorithm's running time.

 Future work focuses on the following two aspects. (1) improve sharding method. For sample space, be sure to make more data transmission local transmission. 
Inevitable data transmission between different machines is dispersed into different transmission inside the rack. Improve transmission concurrency, thereby improving 
transmission efficiency. This requires careful design and implementation of partition strategy of Algorithm~\ref{alg:docInvertShard}. (2) Distributed Parameter Map-Reduce is currently implemented on hadoop. But spark\citep{Spark} stores its data in the form of memory blocks. Its efficiency is easily one or two orders of magnitude higher than hadoop. Distributed Parameter Map-Reduce is composed of a series of map-reduce tasks. These map-reduce tasks can be easily implemented on hadoop. Also, they can be easily implemented on spark. Therefore, Distributed Parameter Map-Reduce algorithm can be easily ported to spark.




\newpage
\vskip 0.2in
\bibliography{sample}

\begin{thebibliography}{13}
\providecommand{\natexlab}[1]{#1}
\providecommand{\url}[1]{\texttt{#1}}
\expandafter\ifx\csname urlstyle\endcsname\relax
  \providecommand{\doi}[1]{doi: #1}\else
  \providecommand{\doi}{doi: \begingroup \urlstyle{rm}\Url}\fi

\bibitem[Ahmed et~al.(2012)Ahmed, Aly, Gonzalez, Narayanamurthy, and
  Smola]{YahooLDA}
Amr Ahmed, Mohamed Aly, Joseph Gonzalez, Shravan Narayanamurthy, and Alexander
  Smola.
\newblock Scalable inference in latent variable models.
\newblock \emph{Proceedings of The 5th ACM International Conference on Web
  Search and Data Mining}, 2012.

\bibitem[Chu et~al.(2014)Chu, Kim, Lin, Yu, Bradski, and Ng]{Chu}
Cheng-Tao Chu, Sang~Kyun Kim, Yi-An Lin, YuanYuan Yu, Gary Bradski, and
  Andrew~Y. Ng.
\newblock Map-reduce for machine learning on multicore.
\newblock \emph{NIPS}, 2014.

\bibitem[Dean and Ghemawat(2004)]{MapReduce}
Jeffrey Dean and Sanjay Ghemawat.
\newblock Mapreduce: Simplified data processing on large clusters.
\newblock \emph{Proceedings of the 6th Operating Systems Design and
  Implementation}, pages 137--150, 2004.

\bibitem[Dean et~al.(2012)Dean, Corrado, Monga, Chen, Devin, Le, Mao, Ranzato,
  Senior, Tucker, Yang, and Ng]{DistBelief}
Jeffrey Dean, Greg~S. Corrado, Rajat Monga, Kai Chen, Matthieu Devin, Quoc~V.
  Le, Mark~Z. Mao, Marc’Aurelio Ranzato, Andrew Senior, Paul Tucker, Ke~Yang,
  and Andrew~Y. Ng.
\newblock Large scale distributed deep networks.
\newblock \emph{NIPS}, 2012.

\bibitem[Feng et~al.(2012)Feng, Kumar, Recht, and Ré]{RDBMS}
Xixuan Feng, Arun Kumar, Benjamin Recht, and Christopher Ré.
\newblock Towards a unified architecture for in-rdbms analytics.
\newblock \emph{SIGMOD}, 2012.

\bibitem[Ghemawat et~al.(2003)Ghemawat, Gobioff, and Leung]{Hdfs}
Sanjay Ghemawat, Howard Gobioff, and Shun-Tak Leung.
\newblock The google file system.
\newblock \emph{Proceedings of the 16th ACM Symposium on Operating System
  Principles}, pages 29--43, 2003.

\bibitem[Ho et~al.(2013)Ho, Cipar, Cui, Kim, Lee, Gibbons, Gibson, Ganger, and
  Xing]{Petumn}
Qirong Ho, James Cipar, Henggang Cui, Jin~Kyu Kim, Seunghak Lee, Phillip~B.
  Gibbons, Garth~A. Gibson, Gregory~R. Ganger, and Eric~P. Xing.
\newblock More effective distributed ml via a stale synchronous parallel
  parameter server.
\newblock \emph{NIPS}, 2013.

\bibitem[Li et~al.(2014)Li, Andersen, Smola, and Yu]{MuLi}
Mu~Li, David~G. Andersen, Alexander Smola, and Kai Yu.
\newblock Communication efficient distributed machine learning with the
  parameter server.
\newblock \emph{NIPS}, 2014.

\bibitem[Low et~al.(1968)Low, Gonzalez, Kyrola, Bickson, Guestrin, and
  Hellerstein]{Graphlab}
Yucheng Low, Joseph Gonzalez, Aapo Kyrola, Danny Bickson, Carlos Guestrin, and
  Joseph~M. Hellerstein.
\newblock Distributed graphlab: A framework for machine learning and data
  mining in the cloud.
\newblock \emph{Proceedings of the VLDB Endowment}, 5\penalty0 (8), 1968.

\bibitem[Nocedal and Wright(2006)]{Optimization}
Jorge Nocedal and Stephen~J. Wright.
\newblock \emph{Numerical Optimization, Second Edition}.
\newblock Springer Verlag, New York, 2006.

\bibitem[Power and Li(2010)]{Piccolo}
Russell Power and Jinyang Li.
\newblock Piccolo: Building fast, distributed programs with partitioned tables.
\newblock \emph{Proceedings of the 6th Operating Systems Design and
  Implementation}, 2010.

\bibitem[Smola and Narayanamurthy(2010)]{ParameterServerTopic}
Alexander Smola and Shravan Narayanamurthy.
\newblock An architecture for parallel topic models.
\newblock \emph{Proceedings of the VLDB Endowment}, 3\penalty0 (1), 2010.

\bibitem[Zaharia et~al.(2012)Zaharia, Chowdhury, Das, Dave, Ma, McCauley,
  Franklin, Shenker, and Stoica]{Spark}
Matei Zaharia, Mosharaf Chowdhury, Tathagata Das, Ankur Dave, Justin Ma, Murphy
  McCauley, Michael~J. Franklin, Scott Shenker, and Ion Stoica.
\newblock Fast and interactive analytics over hadoop data with spark.
\newblock \emph{USENIX}, 37\penalty0 (4):\penalty0 45--51, 2012.

\end{thebibliography}

\newpage
\appendix
\section*{Appendix A.}

\begin{algorithm}[h]
\setlength{\floatsep}{1em}
\setlength{\intextsep}{1em}
   \caption{invertDocumentsSharding(trainInput, docInvertShardOutput)}
   \label{alg:docInvertShard}
\begin{multicols}{2}
\begin{spacing}{0.8}
{\bfseries mapper($longId$, $S^{(i)}$)}
\begin{algorithmic}[1]
   \STATE \{extra\_feature\_info=getExtraInfo()\}
   \STATE $docId=generateRandomId(S^{(i)})$
   \STATE $label=S^{(i)}\rightarrow label$
   \STATE $type='i',num=1$
   \FOR{each $token$ in $S^{(i)}$}
   \STATE $f=token\rightarrow f$
   \STATE $count=token\rightarrow count$ 
   \STATE $units=concat(':',docId,count,label)$
   \STATE $value$=$concat$($'$\hspace{0.3cm}$'$, $type$,$num$, $units$)
   \STATE $fcount=extra_feature_info[f]$
      \FOR{each $f_s$ in $sharding(f,fcount)$}
      \STATE emit($f_s$,$value$)
      \ENDFOR
   \STATE $emit(f, value)$
   \ENDFOR
\end{algorithmic}

{\bfseries combiner(reducer)($f$, $iterator$)}
\begin{algorithmic}[1]
   \STATE  $docUnitsList=\{\}$
   \STATE  $type='i',sum=0$
   \FOR{each $value$ in $iterator$}
   \STATE $sum+=value\rightarrow num$
   \STATE $units=value\rightarrow units$ 
   \STATE $docUnitsList=docUnitsList\cup {units}$
   \ENDFOR
      \FOR{each $f_s$ in $sharding(f,sum)$}
      \STATE $subDocUnits=docUnitsList[f_s]$
      \STATE $num=subDocUnits\rightarrow size$
      \STATE $value$=$concat$($'$\hspace{0.3cm}$'$,$type$,$num$,$subDocUnits$)
      \STATE emit($f_s$,$value$)
      \ENDFOR
\end{algorithmic}
\end{spacing}
\end{multicols}
\end{algorithm}

\vspace{-1em}

\begin{algorithm}[h]
   \caption{invetParamters(docInvertShardOutput, paraInvertOutput)}
   \label{alg:paraSub}
\begin{multicols}{2}
\begin{spacing}{0.8}
{\bfseries mapper($longId$, $paraInvert$)}
\begin{algorithmic}[1]
   \STATE $f_s=paraInvert\rightarrow f_s$
   \STATE $f=f_s\rightarrow f$
   \STATE $emit(f,f_s)$
\end{algorithmic}
{\bfseries combiner(reducer)($f$, $iterator$)}
\begin{algorithmic}[1]
   \STATE  $fsList='',type='e'$
   \FOR{each $f_s$ in $iterator$}
   \STATE $fsList=concat(';',fsList,f_s)$
   \STATE $docUnitsList=docUnitsList\cup\{units\}$
   \ENDFOR
   \STATE $value$=$concat$($'$\hspace{0.3cm}$'$, $type$, $fsList$)
   \STATE $emit(f,value)$
\end{algorithmic}
\end{spacing}
\end{multicols}
\end{algorithm}

\vspace{-1em}
\begin{algorithm}[h]
   \setlength{\floatsep}{1em}
   \caption{distributeParametersSharding(paraValueOutput, paraInvertOutput, paraDistributeShardOutput)}
   \label{alg:paraSubDis}
\begin{multicols}{2}
\begin{spacing}{0.8}
{\bfseries mapper($longId$, $paraInfo$)}
\begin{algorithmic}[1]
   \STATE $f=paraInfo\rightarrow f$
   \STATE $value=paraInfo\rightarrow value$
   \STATE $emit(f,value)$
\end{algorithmic}

{\bfseries reducer($f$, $iterator$)}
\begin{algorithmic}[1]
   \STATE  $fsList='',para='',otype='p'$
   \FOR{each $value$ in $iterator$}
   \STATE $type=value\rightarrow type$
   \IF{$type=='e'$}
   \STATE $fsList=value\rightarrow fsList$
   \ENDIF
   \IF{$type=='p'$}
   \STATE $para=value\rightarrow para$
   \ENDIF
   \ENDFOR
   \STATE $value$=$concat$($'$\hspace{0.3cm}$'$, $otype$, $para$)
   \FOR{each $f_s$ in $fsList$}
   \STATE $emit(f_s,value)$
   \ENDFOR
\end{algorithmic}
\end{spacing}
\end{multicols}
\end{algorithm}

\end{document}